# GENERATION OF A STOCHASTIC BINARY FIELD THAT FITS A GIVEN HETEROGENEITY POWER SPECTRUM

Jiaxuan Li[1] and Yingcai Zheng[1]

[1] Department of Earth and Atmospheric Sciences, University of Houston, Houston, TX

# 1 Abstract

Incomplete binary mixing of two components can form a heterogeneous assemblage in space. The heterogeneity power spectrum of the assemblage can be frequently obtained. However, it is unknown if a stochastic binary field exists to generate the observed spectrum. We propose a novel and powerful constructive procedure for this purpose. The procedure allows us not only to test whether certain binary mixing is feasible but also to tightly constrain the properties of the mixing components and the modal proportion. The method should find wide applications in many branches of geosciences.

## 2   Introduction

Mixing is a general phenomenon widely observed in many dynamic systems. Rather than being too general and vague in the description, we set our discussion in the context of solid Earth sciences. Mixing is a process to either generate or destroy geochemical and geophysical heterogeneities (Allegre & Turcotte, 1986; Bercovici, 2007; Hart, 1988; Hunt & Kellogg, 2001; Kellogg & Turcotte, 1987, 1990; Stixrude & Lithgow-Bertelloni, 2012; van Keken, Hauri, & Ballentine, 2002). Therefore, mapping heterogeneities to infer mixing patterns can provide valuable information in understanding dynamic evolution of the Earth interior.

Among all mixing scenarios, binary mixing (Farley, Natland, & Craig, 1992; Shimizu et al., 2016; Xu, Lithgow-Bertelloni, Stixrude, & Ritsema, 2008) has the fewest mixing components but occupies a special role. For example, the marble-cake model (Allegre & Turcotte, 1986) and the mechanical mixing model of basalt and harzburgite (Xu et al., 2008) provide important frameworks to understand mantle evolution. Small-scale heterogeneity observed from seismic scattering in the lowermost mantle have been interpreted as mantle binary mixing with subducted mid-ocean-ridge basalt (Haugland, Ritsema, van Keken, & Nissen-Meyer, 2018).

When we study binary mixing, it is crucial to distinguish the complete and incomplete mixing. Let us consider mixing two endmember rock types with different seismic velocities. The complete mixing results in a homogenized assemblage whose seismic velocity is constant at the smallest possible spatial scale under investigation. On the other hand, incomplete mixing results in a binary velocity field where at any point we find either one of the two endmember velocities. We can loosely define a "mixing scale" for the random binary field which is characterized by the representative size of spatially contiguous domains of an endmember. In statistical sense, the heterogeneous assemblage has a bulk mean value for some measurable field property (the first moment) and a spatial correlation function for the field fluctuations (the second moment). Usually, the mean value is probed at very low spatial resolution whose scale length is much larger than the mixing scale. Seismic tomography is such an example and can give the large-scale mean seismic velocity in the mantle (Romanowicz, 2003). However, knowing only the mean velocity has limited power in understanding binary mixing at the "mixing scale" because there are many possible mixing endmembers that can yield the same mean value. On the other hand, if the correlation function of the velocity fluctuation can also be obtained, the mixing can be much better understood. Seismic scattering can determine the power spectrum (hence spatial correlation function) of the random heterogeneity using transmission fluctuations (Aki, 1973; Chen & Aki, 1991; Chernov, 1960; Cormier, 2000; Wu & Flatté, 1990; Zheng & Wu, 2005; Zheng & Wu, 2008) or seismic envelopes (Sato & Fehler, 1998). The power spectrum is defined as the square of the Fourier amplitude spectrum and therefore does not have the phase information. It is usually straightforward to find a random realization of a continuous (not binary) field that can generate the observed heterogeneity power spectrum (Muller, Roth, & Korn, 1992; Zheng & Wu, 2008). However, <u>the question is: can we find a stochastic discrete binary velocity field such that it can produce the heterogeneity power spectrum?</u>

# 3 Method and Examples

## 3.1 Definition of heterogeneity power spectrum and two-point correlation function

Suppose the heterogenous seismic velocity field in the mantle is $v(\mathbf{x})$. The background velocity is $v_0$. Then the fluctuation of the velocity field (Zheng & Wu, 2008) could be defined as:

$$\varepsilon(\mathbf{x}) = \frac{1}{2}\left[\frac{v_0^2}{v^2(\mathbf{x})} - 1\right] \approx \frac{v(\mathbf{x})}{v_0} - 1 = \frac{\delta v(\mathbf{x})}{v_0} \ . \tag{1}$$

The two-point correlation function, $B(\mathbf{r})$, of the fluctuation field is defined as

$$B(\mathbf{r}) = \overline{\varepsilon(\mathbf{x})\varepsilon(\mathbf{x}+\mathbf{r})} \ , \tag{2}$$

where the overbar means averaging the product over all possible locations $\mathbf{x}$ for a given a correlation lag $\mathbf{r}$. The heterogeneity spectrum of the fluctuation field is defined as

$$F(\mathbf{k}) = \iiint \varepsilon(\mathbf{x}) e^{-i\mathbf{k}\cdot\mathbf{x}} d^3\mathbf{x} \ , \tag{3}$$

which is usually a complex number. The heterogeneity power spectrum $P(\mathbf{k})$ is just the Fourier transform of the fluctuation correlation $B(\mathbf{r})$:

$$P(\mathbf{k}) = \iiint B(\mathbf{r}) e^{-i\mathbf{k}\cdot\mathbf{r}} d^3\mathbf{r} = |F(\mathbf{k})|^2 \ . \tag{4}$$

Therefore, knowing the correlation function can enable us to obtain the power spectrum, and vice versa.

We further assume the binary mixture is a statistically isotropic random medium, which means that the heterogeneity correlation function does not depend on the orientation and we have $B(r) = B(\mathbf{r})$, where $r$ is the length of the vector $\mathbf{r}$.

It is easy to see that the two-point correlation function of the velocity field $C(r) = C(\mathbf{r}) = \overline{v(\mathbf{x})v(\mathbf{x}+\mathbf{r})}$ has the following relationship with the fluctuation correlation function of the random field $B(r)$ using equation (1):

$$C(r) = v_0^2 \left[1 + B(r)\right] . \tag{5}$$

### 3.2 Problem statement and binary field construction

Our goal is to construct a stochastic binary velocity field $V^{(B)}(\mathbf{x})$, whose spatial correlation function is the same as $C(r)$.

To achieve this goal, we applied a two-step reconstruction method.

In the first step, we will determine a viable combination of the two endmember velocities and the volume fraction of each component.

In the second step, we will apply the simulated annealing (SA) optimization method (Yeong & Torquato, 1998) to find a stochastic realization of a binary field that could yield the correlation function $C(r)$.

Step1: Determination velocities and volume fractions of the mixing endmembers

Suppose the two mixing components have two discrete velocities: $v_a$ and $v_b$ for phase-a and phase-b, respectively. The binary velocity field $V^{(B)}$ takes a velocity either $v_a$ or $v_b$ at any location **x** and is sought to fit the following constraint

$$C(r) = \overline{V^{(B)}(\mathbf{x})V^{(B)}(\mathbf{x}+\mathbf{r})} \ . \tag{6}$$

It is more convenient for us to use an indicator function defined as:

$$I^{(a)}(\mathbf{x}) = \begin{cases} 1 & \text{if } \mathbf{x} \in \text{phase a,} \\ 0 & \text{if } \mathbf{x} \notin \text{phase a.} \end{cases} \tag{7}$$

The correlation function of the indicator field for phase-a is defined as:

$$S^{(a)}(r) = S^{(a)}(\mathbf{r}) = \overline{I^{(a)}(\mathbf{x})I^{(a)}(\mathbf{x}+\mathbf{r})} \ . \tag{8}$$

It is straightforward to show the following relationship between $C(r)$ and $S^{(a)}(r)$:

$$C(r) = \frac{S^{(a)}(r) - a_1\phi_a - a_2}{a_3} \ , \tag{9}$$

where

$$a_1 = \frac{2v_b}{v_a - v_b}, \quad a_2 = -\frac{v_b^2}{v_a - v_b}, \quad a_3 = \frac{1}{(v_a - v_b)^2} \ . \tag{10}$$

The bulk volume fraction of phase-a, is given by the correlation function $I^{(a)}(\mathbf{x})$ at the zero lag,

$$S^{(a)}(0) = \phi_a \ . \tag{11}$$

On the other hand, when the lag $R$ is large enough, we have

$$S^{(a)}(R) = \phi_a^2 \ . \tag{12}$$

To see the validity of equation (12), we use the following rationale. At any point in the binary field, the probability to find phase-a is $\phi_a$. The probability of finding phase-a simultaneously at two locations (see equation (8)) at a distance $R$ apart is $\phi_a^2$ when $R$ is large. The reason we need a large $R$ is that the field is decorrelated at large distances.

Plugging (5), (11), and (12) into equation (9), we can obtain the relationship $v_a$, $v_b$, and $\phi_a$:

$$v_a(\phi_a) = v_0 \left[ \sqrt{1+B(R)} + \sqrt{\frac{1-\phi_a}{\phi_a}} \sqrt{B(0)-B(R)} \right],$$
$$v_b(\phi_a) = v_0 \left[ \sqrt{1+B(R)} - \sqrt{\frac{\phi_a}{1-\phi_a}} \sqrt{B(0)-B(R)} \right].$$
(13)

Here, $B(0)$ and $B(R)$ can be taken as known from the observed correlation function $B(r)$ and $v_0$ is the background velocity. Given a volume fraction of phase-a, $\phi_a$, we can determine the two endmember velocities $v_a$ and $v_b$ using equation (13). We still need to show that there is a stochastic binary field whose correlation function can fit the observed correlation function.

Step2: Determination of the stochastic structure of binary-mixing field

Because there is a one-to-one correspondence between the fluctuation correlation $B(r)$ and the indicator correlation function $S^{(a)}(r)$ using equation (9), we choose to find a binary field with a correlation that fits $S^{(a)}(r)$.

We will apply an iterative simulated annealing (SA) method (Kirkpatrick, Gelatt, & Vecchi, 1983; Yeong & Torquato, 1998) for this purpose. We start from some random binary indicator field. We then update the binary indicator field in order to reach the final stochastic field with $S^{(a)}(r)$. In each SA step, we propose to minimize the misfit $E$ defined in the least-squares sense:

$$E = \sum_i \left[ S^{SA}(r_i) - S^{(a)}(r_i) \right]^2, \qquad (14)$$

where $r_i$ denotes the $i^{th}$ correlation lag, and $S^{SA}$ is the correlation function for the binary indicator field at that SA step. For each lag $r_i$, we calculate the average of all corresponding two-point products (Yeong & Torquato, 1998).

In each SA step, we swap two points randomly chosen from the indicator field. In this way, the bulk volume fraction of each phase will be automatically conserved. We then calculate the new misfit $E'$ and we can compute the change in the misfit $dE = E' - E$ before and after each swap. If $dE \leq 0$, we will accept the swap with probability 1. If $dE > 0$, the swap is accepted according to the probability, $p(dE)$, defined as

$$p(dE) = \begin{cases} 1 & dE \leq 0 \\ \exp(-dE/T) & dE > 0 \end{cases}. \qquad (15)$$

The reason we may still accept the swap even the misfit increases is to avoid being trapped in the local minimum in iteration.

The starting value of $T$ (usually called "melting temperature" in SA) should be a value large enough to make the acceptance ratio in the initial several hundred swaps greater than 80%. If the acceptance ratio is less than 80%, the initial $T$ should be doubled. After the initial "melting temperature" $T$ is determined, the "cooling process" starts, the parameter $T$ decreases at a constant rate, say 0.95, after certain number of swaps (Kirkpatrick, 1984). The iteration terminates when the misfit $E$ given by equation (14) is less than some given tolerance value such that the misfit is minimized in the sense of the least-squares.

## Examples

We provide two examples to show how to construct a stochastic binary field whose correlation/spectrum can fit a given correlation/spectrum function. We will test two types of 2D correlation functions for the fluctuations, Gaussian and von Karman types. For the Gaussian type, its correlation function is:

$$B(r) = \chi^2 e^{-(r/r_0)^2} . \tag{16}$$

For the von Karman type, its correlation function is:

$$B(r) = \frac{2^{1-\nu}}{\Gamma(\nu)} \chi^2 \left(\frac{r}{r_0}\right)^{\nu} K_{\nu}\left(\frac{r}{r_0}\right) . \tag{17}$$

In equations (16) and (17), $\chi$ is the root mean square (rms) of the random fluctuation field. $r$ is the correlation lag. $r_0$ is a length scale for the heterogeneities. In equation (17), $\Gamma$ is the gamma function, $K_{\nu}$ is the modified Bessel function, $\nu$ is the Hurst number ($0 < \nu < 1$) that controls the amount of small-scale heterogeneities relative to the large-scale ones.

## Gaussian Type:

For the Gaussian type, we set the rms velocity $\chi = 0.03$ km/s, $r_0 = 0.2 km$ to obtain the correlation function $B(r)$ (Figure 1a). Suppose the corresponding fluctuation field has a dimension of 200×200 pixels, and the pixel size is the same along the x direction ($\Delta x$) and the z direction ($\Delta z$), i.e., $\Delta x = \Delta z = 0.05\ km$. Given a background velocity $v_0$ ( say, $v_0 = 8km/s$ ), a random realization of the velocity field is $v(x,z)$ (Figure 2e) if we were to find a continuous random field that can fit the correlation function $B(r)$. However, our goal is to find a stochastic binary field that genearte the same $B(r)$.

With $B(0), B(R)$, and $v_0$, the relationship of the binary velocities $v_a$ (red dashed line) and $v_b$ (blue dash-dotted line) with respect of the volume fraction, $\phi_a$, can be determined (Figure 1b) (equation (13)). For instance, by fixing $\phi_a = 0.3$, we can determine the velocities of the two mixing endmembers, $v_a = 8.36 km/s$ and $v_b = 7.83 km/s$, respectively. If these velocities do not correspond to properties of real geological components, we can continue to vary $\phi_a$ until the two velocities make geological sense. Therefore, this is a powerful method to constrain possible $v_a$, $v_b$, and $\phi_a$ if the binary mixing scenario is assumed.

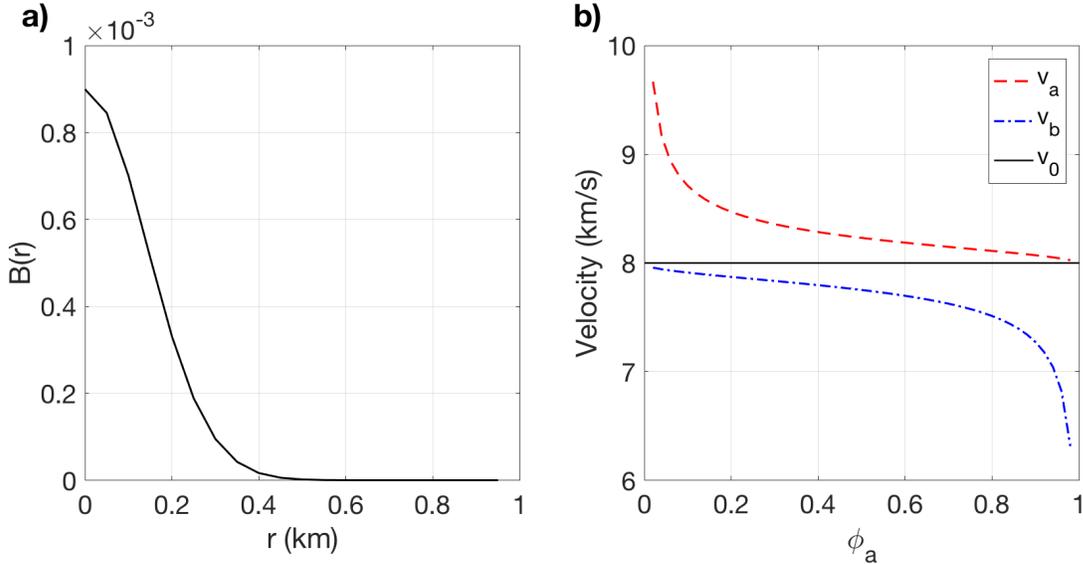

Figure 1. a) A Gaussian correlation function. b) Relationship of binary velocities $v_a$ (red dashed line) and $v_b$ (blue dash-dotted line) with the volume fraction of phase-$a$ $\phi_a$ for this Gaussian type. The black solid line represents the background velocity $v_0$.

Next, we will find a random binary field that corresponds to $B(r)$. To do this, we choose to work on the binary indicator correlation function $S^{(a)}(r)$ which has a one-to-one correspondence with $B(r)$ (Figure 2a) (also see equations (5) and (9)).

We create an initial random binary field model with the volume fraction of phase-$a$ $\phi_a = 0.3$ (Figure 2c). The correlation function of the corresponding indicator field is far from the observed $S^{(a)}(r)$ (Figure 2a). Then we apply the SA iteration. At each SA step, we compute the correlation function $S^{SA}(r)$ and its misfit (Figure 2b). After 2 million SA steps, we achieved a correlation $S^{SA}(r)$ that fits the true correlation function $S^{(a)}(r)$ well (Figure 2a) and the misfit is decreased to an acceptable level (Figure 2b).

The final stochastic binary field (Figure 2d) from the SA method shows many spatially contiguous domains or "blobs". These blobs have similar spatial scales which is characteristic to the single-scale Gaussian random field.

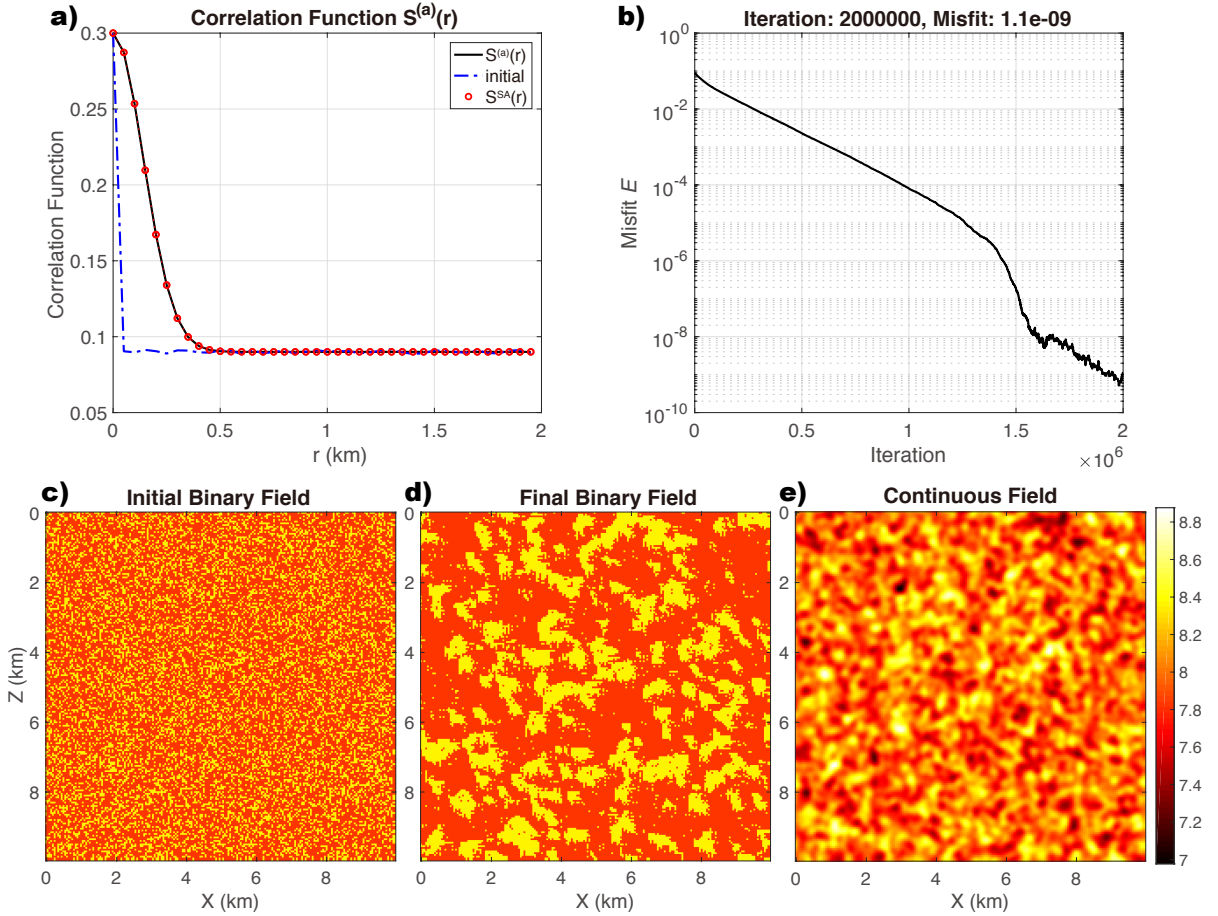

Figure 2. Gaussian type binary field. **a)** Comparison of the observed correlation function ($S^{(a)}$, black solid line), initial correlation function (blue dash-dotted line), and final correlation function of the indicator field after 2 million simulated annealing iterations ($S^{SA}$, red circles); **b)** Correlation misfit, $E$ (equation (14)), with the iteration number; **c)** Initial random binary velocity with volume fraction $\phi_a = 0.3$ for phase-$a$. **d)** Final binary velocity field $V^{(B)}(x,z)$ using SA. **e)** One realization of the continuous velocity field given the observed $B(r)$ and background velocity $v_0$. In **c)** and **d)**, the yellow region denotes phase-a and red region denotes phase-b.

von Karman Type:

We set the rms velocity $\chi = 0.03 \; km/s$, $r_0 = 0.4 km$, and $\nu = 0.2$ to obtain one von Karman type correlation function $B(r)$ (Figure 3 a). Suppose the corresponding fluctuation field has a dimension 400×400 pixels, each pixel is a grid with dimension $0.05km \times 0.05km$. We can then obtain one random realization of the continuous velocity field (Figure 4e) given background velocity $v_0 = 8km/s$.

Similar to the Gaussian type, we can determine the relationship of the binary velocities $v_a$ (red dashed line) and $v_b$ (blue dash-dotted line) with volume fraction $\phi_a$ (Figure 3b).

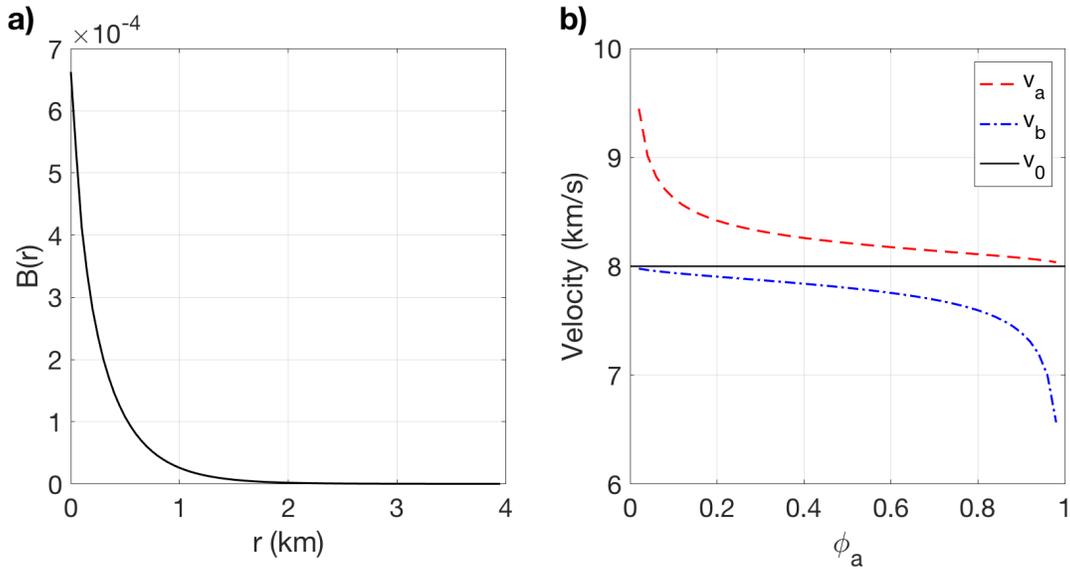

Figure 3. **a)** von Karman correlation function. **b)** Relationship of binary velocities $v_a$ (red dashed line) and $v_b$ (blue dash-dotted line) with the volume fraction of phase-$a$ $\phi_a$ for the von Karman type medium. The black solid line represents the background velocity $v_0$.

Given a volume fraction $\phi_a = 0.3$, we obtain $v_a = 8.33km/s$ and $v_b = 7.87km/s$ using equation (13). The structure of the binary field could be obtained in a way similar to the one for the Gaussian type after 3 million iterations. The $S^{SA}(r)$ that fits the observed correlation function $S^{(a)}(r)$ well and the misfit was reduced to the level of $10^{-10}$ (Figure 4b). The final SA binary velocity field (Figure 4d) shows that the spatial distribution of phase-a is less contiguous compared to the Gaussian case (Figure 2d). There are multiple sizes for the contiguous domains dispersed throughout the model, which are characteristic for the multi-scale von Karman random medium.

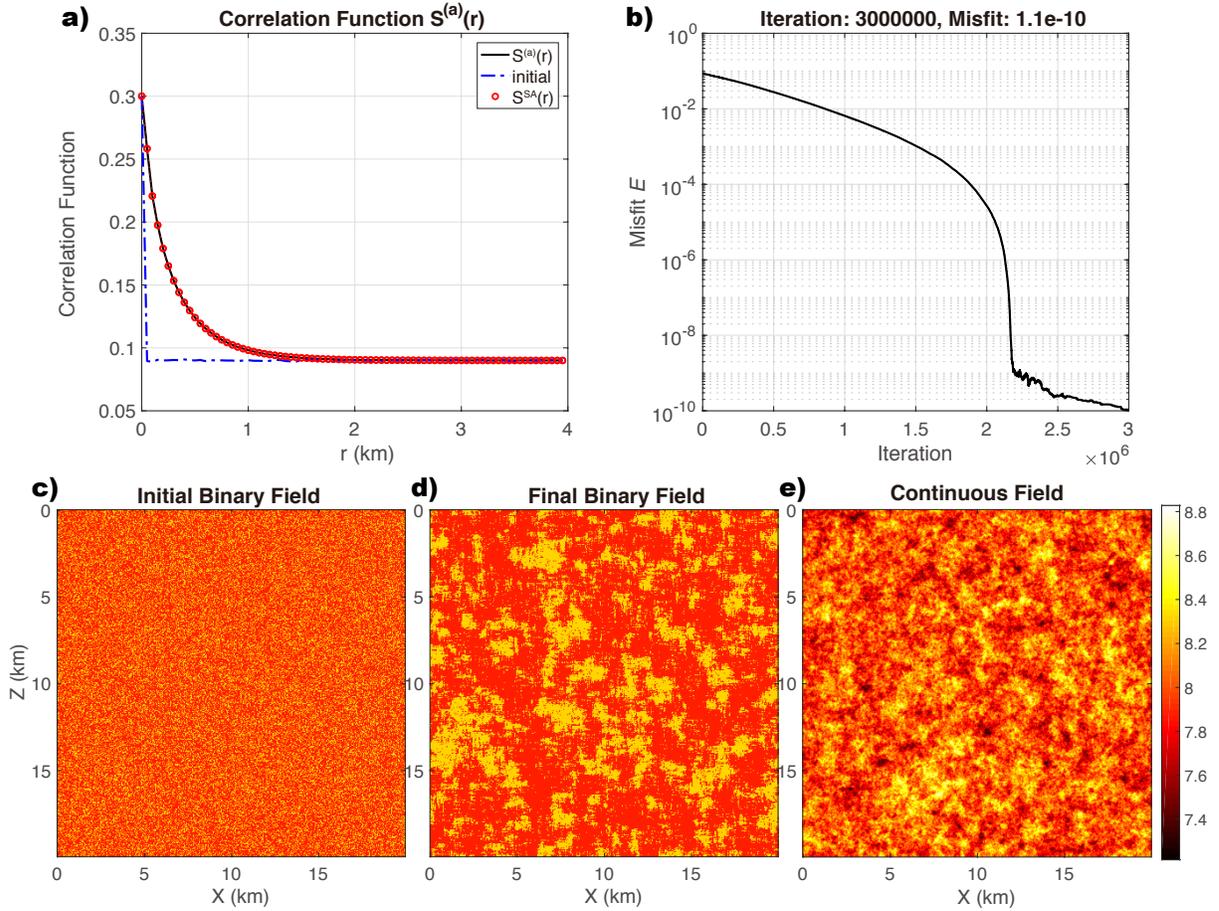

Figure 4. von Karman binary field. **a)** Comparison of the observed correlation function (black solid line), initial correlation function (blue dash-dotted line), and final SA correlation function (red circles) of binary indicator field after 3 million iterations. **b)** Misfit $E$ (equation (14)) with the iteration number. **c)** Initial random binary velocity field with volume fraction $\phi_a = 0.3$. **d)** Final binary-mixing velocity field $V^{(B)}(x,z)$ after 3 million steps of SA. **e)** One random realization of continuous velocity field given observed $B(r)$ and background velocity $v_0$. In **c)** and **d)**, the yellow region denotes phase-a and red region denotes phase-b.

### 3.3 Discussions and Conclusions

Our iterative simulated annealing approach provides a new way to test the feasibility of geological binary-mixing models such as binary models for the mantle (Farley et al., 1992; Shimizu et al., 2016; Xu et al., 2008). If some region in the mantle is composed of two types of rocks such as basalt and harzburgite, we can then constrain the velocities and volume fraction of basalt and harzburgite using the observed heterogeneity spectrum. This can be potentially useful in understanding mantle evolution and recycling of slab materials.

In conclusion, we find a novel constructive approach using simulated annealing to construct a random binary field that fits both the mean value and the correlation function (or heterogeneity power spectrum) of random heterogeneities. This new method enables us to obtain a tight relationship of the binary velocities and the volume fraction of the binary endmembers. Spatial patterns of the random field can be used to infer dynamic styles of mixing. The method is general and is a new powerful analysis tool for understanding binary mixing in a variety of settings.

## 4 Acknowledgements

Data and MATLAB codes that support the results of this study will be published as online supplement materials and assistance will be available upon request to the authors. This work is supported by NSF EAR-1621878. The authors declare that they have no competing interests.

References

Aki, K. (1973). Scattering of P waves under the Montana Lasa. *Journal of Geophysical Research, 78*(8), 1334-1346.

Allegre, C. J., & Turcotte, D. L. (1986). Implications of a two-component marble-cake mantle. *Nature, 323*(6084), 123-127.

Bercovici, D. (2007). Mantle Dynamics Past, Present, and Future: An Introduction and Overview. In S. Gerald (Ed.), *Treatise on Geophysics* (pp. 1-30). Amsterdam: Elsevier.

Chen, X., & Aki, K. (1991). General coherence functions for amplitude and phase fluctuations in a randomly heterogeneous medium. *Geophysical Journal International, 105*(1), 155-162.

Chernov, L. A. (1960). *Wave Propagation in a Random Medium*. New York: McGraw-Hill.

Cormier, V. F. (2000). D " as a transition in the heterogeneity spectrum of the lowermost mantle. *Journal of Geophysical Research-Solid Earth, 105*(B7), 16193-16205.

Farley, K. A., Natland, J. H., & Craig, H. (1992). BINARY MIXING OF ENRICHED AND UNDEGASSED (PRIMITIVE-QUESTIONABLE) MANTLE COMPONENTS (HE, SR, ND, PB) IN SAMOAN LAVAS. *Earth and Planetary Science Letters, 111*(1), 183-199. doi:10.1016/0012-821x(92)90178-x

Hart, S. R. (1988). HETEROGENEOUS MANTLE DOMAINS - SIGNATURES, GENESIS AND MIXING CHRONOLOGIES. *Earth and Planetary Science Letters, 90*(3), 273-296. doi:10.1016/0012-821x(88)90131-8

Haugland, S. M., Ritsema, J., van Keken, P. E., & Nissen-Meyer, T. (2018). Analysis of PKP scattering using mantle mixing simulations and axisymmetric 3D waveforms. *Physics of the Earth and Planetary Interiors, 276*, 226-233. doi:https://doi.org/10.1016/j.pepi.2017.04.001

Hunt, D. L., & Kellogg, L. H. (2001). Quantifying mixing and age variations of heterogeneities in models of mantle convection: Role of depth-dependent viscosity. *Journal of Geophysical Research-Solid Earth, 106*(B4), 6747-6759. doi:10.1029/2000jb900261

Kellogg, L. H., & Turcotte, D. L. (1987). HOMOGENIZATION OF THE MANTLE BY CONVECTIVE MIXING AND DIFFUSION. *Earth and Planetary Science Letters, 81*(4), 371-378. doi:10.1016/0012-821x(87)90124-5

Kellogg, L. H., & Turcotte, D. L. (1990). Mixing and the Distribution of Heterogeneities in a Chaotically Convecting Mantle. *Journal of Geophysical Research, 95*(B1), 421–432.

Kirkpatrick, S. (1984). Optimization by simulated annealing: Quantitative studies. *Journal of Statistical Physics, 34*(5), 975-986. doi:10.1007/bf01009452

Kirkpatrick, S., Gelatt, C. D., & Vecchi, M. P. (1983). Optimization by Simulated Annealing. *Science, 220*(4598), 671-680. doi:10.1126/science.220.4598.671

Muller, G., Roth, M., & Korn, M. (1992). SEISMIC-WAVE TRAVEL-TIMES IN RANDOM-MEDIA. *Geophysical Journal International, 110*(1), 29-41. doi:10.1111/j.1365-246X.1992.tb00710.x

Romanowicz, B. (2003). Global mantle tomography: Progress status in the past 10 years. *Annual Review of Earth and Planetary Sciences, 31*, 303-328. doi:10.1146/annurev.earth.31.091602.113555

Sato, H., & Fehler, M. (1998). *Seismic Wave Propagation and Scattering in the Heterogeneous Earth*. New York: Springer-Verlag.

Shimizu, K., Saal, A. E., Myers, C. E., Nagle, A. N., Hauri, E. H., Forsyth, D. W., . . . Niu, Y. L. (2016). Two-component mantle melting-mixing model for the generation of mid-ocean ridge basalts: Implications for the volatile content of the Pacific upper mantle. *Geochimica Et Cosmochimica Acta, 176*, 44-80. doi:10.1016/j.gea.2015.10.033

Stixrude, L., & Lithgow-Bertelloni, C. (2012). Geophysics of Chemical Heterogeneity in the Mantle. *Annual Review of Earth and Planetary Sciences, 40*(1), 569-595. doi:10.1146/annurev.earth.36.031207.124244

van Keken, P. E., Hauri, E. H., & Ballentine, C. J. (2002). MANTLE MIXING: The Generation, Preservation, and Destruction of Chemical Heterogeneity. *Annual Review of Earth and Planetary Sciences, 30*(1), 493-525. doi:10.1146/annurev.earth.30.091201.141236

Wu, R.-S., & Flatté, S. M. (1990). Transmission fluctuations across an array and heterogeneities in the crust and upper mantle, in "Seismic Wave Scattering and Attenuation" ed. by Wu and Aki,. *Pure and Applied Geophysics, 132*(1-2), 175-196.

Xu, W. B., Lithgow-Bertelloni, C., Stixrude, L., & Ritsema, J. (2008). The effect of bulk composition and temperature on mantle seismic structure. *Earth and Planetary Science Letters, 275*(1-2), 70-79. doi:Doi 10.1016/J.Epsl.2008.08.012

Yeong, C. L. Y., & Torquato, S. (1998). Reconstructing random media. *Physical Review E, 57*(1), 495-506. doi:10.1103/PhysRevE.57.495

Zheng, Y., & Wu, R.-S. (2005). Measurement of phase fluctuations for transmitted waves in random media. *Geophysical Research Letters, 32*(14), L14314. doi:10.1029/2005gl023179

Zheng, Y., & Wu, R. S. (2008). Theory of Transmission Fluctuations in a Depth-dependent Background Medium. In H. Sato & M. Fehler (Eds.), *Earth Heterogeneity and Scattering Effects on Seismic Waves* (Vol. 50, pp. 21-41).